# Breast Cancer Image Classification Method Based on Deep Transfer Learning

Weimin Wang *

The Hong Kong University of Science and Technology, Hong Kong, China, wangwaynemin@gmail.com

Yufeng Li

University of Southampton, Southampton, UK, liyufeng0913@gmail.com

Xu Yan

Trine University, Phoenix, Arizona, US, yancontiue@gmail.com

Mingxuan Xiao

Southwest Jiaotong University, Nanjing, China, 553556963albert@gmail.com

Min Gao

Trine University, Allen Park, Michigan, US, mingao4460@gmail.com

**Abstract**

To address the issues of limited samples, time-consuming feature design, and low accuracy in detection and classification of breast cancer pathological images, a breast cancer image classification model algorithm combining deep learning and transfer learning is proposed. This algorithm is based on the Densely Connected Convolutional Networks (DenseNet) structure of deep neural networks, and constructs a network model by introducing attention mechanisms, and trains the enhanced dataset using multi-level transfer learning. Experimental results demonstrate that the algorithm achieves an efficiency of over 84.0% in the test set, with a significantly improved classification accuracy compared to previous models, making it applicable to medical breast cancer detection tasks.

## 1 INTRODUCTION

Breast cancer is one of the most significant malignant tumors affecting women's health worldwide, ranking first in female malignancy mortality. In China, the incidence and mortality rates of breast cancer among women account for 11.2% and 9.2% of the global total, respectively, placing it at the forefront worldwide. Currently, the etiology of tumors remains unclear, and effective screening methods are lacking, leading to most patients being diagnosed at intermediate or late stages, missing the optimal treatment window. Therefore, early diagnosis and treatment of breast cancer play a crucial role in improving women's health. Currently, clinical diagnosis of breast cancer typically relies on physicians' observation of pathological images of breast cancer tissue, which is not only time-consuming but also highly dependent on the physician's expertise and experience, thus subjective. Yulu, GONG et al. [1] summarized that deep learning technology, with its robust capability for feature learning and representation, has emerged as an invaluable aid for physicians, markedly enhancing diagnostic accuracy and efficiency. And based on [2] and [3] already achieved high accuracy on classification in Alzheimer diagnosis. Therefore, achieving medical image classification based on breast cancer pathological images is of great significance in helping physicians improve diagnostic efficiency.

Currently, classification of breast cancer pathological images can be divided into two main categories: one is medical image classification based on manual feature extraction and traditional machine learning algorithms. Spanhol et al. [4] released the BreakHis dataset of breast cancer tissue pathological images, studying six manually extracted texture features and combining four classifiers including SVM (Support Vector Machine) to distinguish between benign and malignant tumors. Ojalat [5] proposed the Local Binary Pattern (LBP), which is an operator used to describe image texture features, thereby improving the classification of texture features to differentiate between benign and malignant breast cancer images. Rezazadeh A [6] proposed an interpretable machine learning method for diagnosing breast cancer based on ultrasound images. By extracting first and second-order texture features from ultrasound images, they constructed a probability ensemble of decision tree classifiers. Each decision tree learned to classify input ultrasound images by learning a set of robust decision thresholds for image texture features, and decomposing the learned decision trees to explain the decision paths of the model. The results showed that the proposed framework achieved high predictive performance while being interpretable.

Traditional machine learning methods suffer from a lack of experienced pathologists to annotate image features, and the extraction and selection of classification features often consume a considerable amount of time and effort, ultimately resulting in suboptimal accuracy. In contrast to traditional machine learning classification algorithms, deep learning can automatically learn features from images, thereby avoiding the complexity and limitations of manually extracting features in traditional algorithms. Moreover, deep learning is widely applied in various fields such as natural language processing, object recognition, and image classification, laying the foundation for its application in breast cancer pathological images.

For instance, Pawer M M et al. [7] proposed a multi-scale multi-channel feature network (MuSCFNet) combining ResNet (Residual Network) with attention mechanisms and employing a knowledge sharing strategy, achieving a classification accuracy of up to 98.85% in the binary classification task of breast pathological tissue images. Kavitha T et al. [8] introduced a breast cancer diagnostic model based on optimal multi-level threshold segmentation and capsule networks (OMLTS-DLCN), achieving accuracies of 98.50% on the Mini-MIAS (Mammographic Image Analysis Society Mini-database) dataset and 97.55% on the DDSM (Digital Database for Screening Mammography) dataset.

Recent studies have demonstrated that using deep learning methods to classify breast cancer tissue pathological images can significantly improve classification accuracy, thereby assisting physicians in diagnosis and ensuring timely treatment for patients. However, deep learning is highly dependent on data; the larger the dataset, the more helpful it is for the network's classification accuracy. However, in reality, acquiring large medical image datasets is challenging. Additionally, increasing the depth of neural networks in deep learning does not necessarily improve classification accuracy; instead, it may lead to performance degradation. To address these limitations, this study proposes a breast cancer pathological image classification method based on deep transfer learning.

## 2 DEEP TRANSFER LEARNING

### 2.1 Transfer Learning

In medical image classification, data dependency is a crucial concern. Due to the uniqueness of medical images, available medical data for research purposes is often limited and scarce. However, training deep neural network structures such as Convolutional Neural Networks (CNNs) with a small amount of data may lead to overfitting, thereby affecting experimental results. Therefore, in this study, transfer learning is introduced.

Transfer learning is a process of leveraging knowledge learned from one domain (source domain) to aid learning in another domain (target domain) by exploiting similarities between data, tasks, or models. We can utilize transfer learning to transfer the trained model parameters to a new model to aid its training. Specifically, by training a network on a very large dataset and then transferring its pre-trained learning parameters, especially weights, to the target network model, we can provide the target model with powerful feature extraction capabilities, reducing computation time and storage requirements. Transfer learning has been widely applied in medical imaging, showing significant efficacy in terms of accuracy, training time, and error rates.

In transfer learning, fine-tuning parameters addresses the issue of mismatch between pre-trained neural network model feature parameters and the task in the target domain, which is the most crucial step. In this study, since the target domain dataset is relatively small and significantly different from the source domain dataset in terms of image characteristics, the primary approach adopted is to freeze the model for fine-tuning [9].

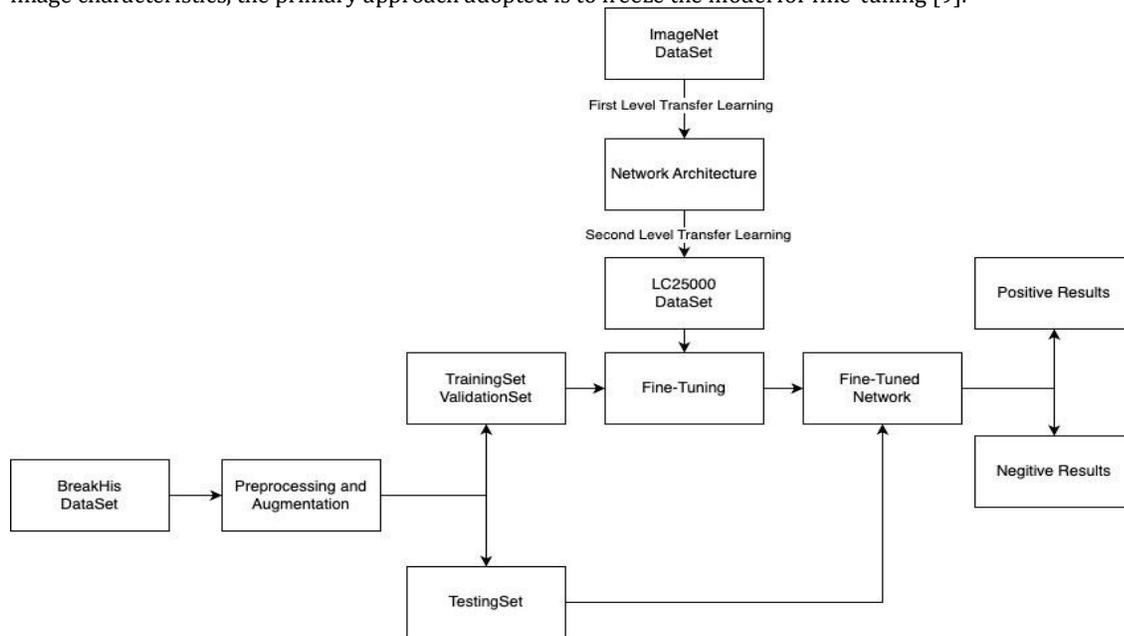

Figure 1. Flow chart of deep transfer learning.

## 2.2 Deep Transfer Learning

In this study, a deep transfer learning method is proposed. Considering that when using the ImageNet dataset as the source domain for transfer learning, the large quantity and non-relevance to cancer of ImageNet data compared to the limited amount of cancer-related data can lead to a significant dissimilarity between the domains, resulting in negative transfer phenomenon and affecting classification accuracy.

In simple terms, in convolutional neural networks, shallow layers are responsible for extracting basic features, while deep layers extract abstract features. Since basic features in images are universal, the shallow layers are pre-trained using the ImageNet dataset. After training the shallow layers, the deep layers are further pre-trained using other cancer-related datasets to exploit the distinct properties of shallow and deep layers for transfer learning, thereby avoiding negative transfer caused by low dataset similarity.

This study proposes a breast cancer medical image classification algorithm based on deep transfer learning. Firstly, the Densely Connected Convolutional Networks (DenseNet) is selected as the network architecture for this study and is improved by integrating attention mechanisms to enhance its performance. Then, the improved DenseNet is subjected to the first transfer learning using the ImageNet natural image dataset. After the first transfer learning, the network is further fine-tuned through the second transfer learning using the LC2500 lung cancer dataset. Subsequently, the network is trained using the preprocessed and augmented BreakHis breast cancer dataset via fine-tuning. Finally, the trained network model's classification accuracy is evaluated using the test set. The workflow is illustrated in Figure 1 above.

## 3 DATA COLLECTION AND PREPROCESSING

### 3.1 The BreakHis dataset

Spanhal has publicly released a breast cancer tissue pathological image dataset named BreakHis. The dataset comprises 82 cases, totaling 7,909 breast cancer images, including 2,480 benign images and 5,429 malignant images. Each image has dimensions of 700 × 460 pixels, with a PNG format and 3 channels. These images are obtained through biopsy, where breast cancer tissue slices are placed under a microscope and imaged at different magnification levels. Finally, experienced pathologists diagnose these images, classifying them as benign or malignant. Figure 2 illustrates pathological images at four magnification levels, and Table 1 provides statistics on the dataset.

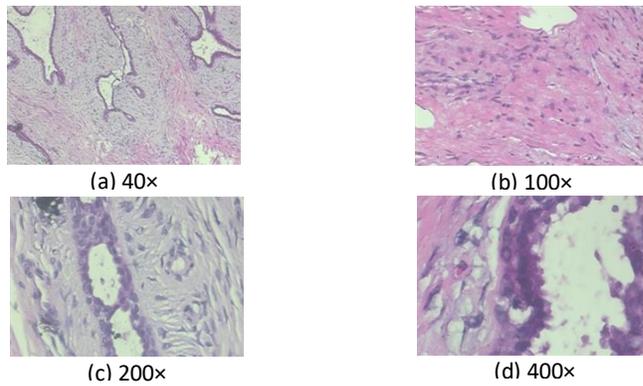

(a) 40×  (b) 100×  
(c) 200×  (d) 400×  

Figure 2. Image of breast cancer pathology at four magnification.

Table 1. Data set data statistics.

| Magnification | Benign | Malignant | Total |
| --- | --- | --- | --- |
| 40× | 625 | 1370 | 1995 |
| 100× | 644 | 1437 | 2081 |
| 200× | 623 | 1390 | 2013 |
| 400× | 588 | 1232 | 1820 |
| Total | 2480 | 5429 | 7909 |
| Cases | 24 | 58 | 82 |

### 3.2 Image Preprocessing and Data Augmentation

In pathological tissue images, to mitigate the impact of noise on experimental classification accuracy and to simplify data, thereby enhancing the training speed of the network model, preprocessing of the dataset is required. In the BreakHis dataset, pathological images undergo staining, and the color variations may significantly affect the final classification results [10]. Therefore, in this study, the original images are first subjected to color normalization preprocessing. Subsequently, data augmentation is applied to the BreakHis dataset to increase the dataset size and reduce overfitting during network model training. Common data augmentation techniques include flipping, rotation, scaling, cropping, translation, and adjusting colors and contrasts. In this study, the dataset is augmented by rotating 90 degrees, 180 degrees, 270 degrees, horizontally flipping, and vertically flipping, resulting in a dataset size five times larger than the original dataset. Part of the preprocessed dataset is shown in Figure 3, where Figure 3a represents the original image, Figure 3b represents the image after color normalization, and Figure 3c represents the image after rotating 270 degrees.

## 4 DEEP LEARNING MODEL

### 4.1 DenseNet

The DenseNet network employs a unique connection scheme known as dense connectivity, where each layer in the dense block utilizes feature maps generated by all preceding layers in the block as its input, and subsequently, the feature maps generated by the layer are used as inputs for all subsequent layers [11], [12]. Through this dense connectivity, DenseNet networks efficiently utilize feature information from multiple preceding layers, thereby reducing parameters while improving network efficiency.

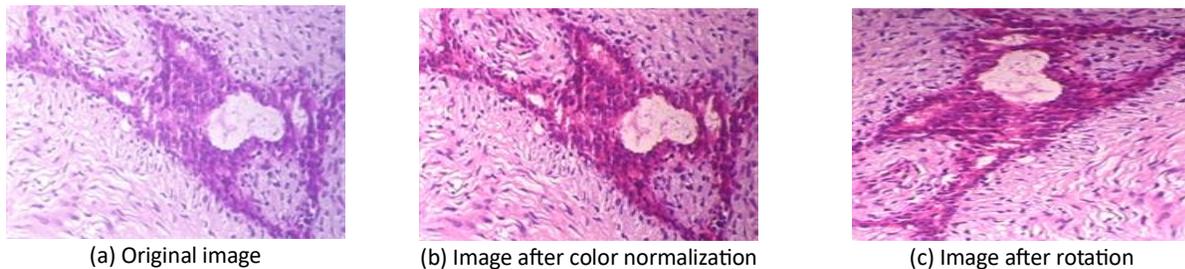

(a) Original image　　　　　　　(b) Image after color normalization　　　　　　　(c) Image after rotation

Figure 3. Comparison of the original and pre-processed images.

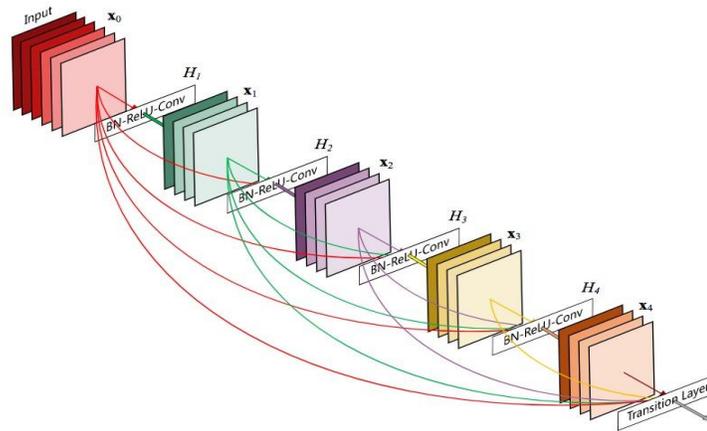

Figure 4. DenseNet network structure.

The DenseNet architecture is characterized by its composition of Dense Blocks and Transition Layers. Dense Blocks contain several layers which are densely connected, allowing feature maps from preceding layers to be concatenated within each block. This dense connectivity facilitates the flow of information and promotes feature reuse throughout the network. Transition layers are inserted between dense blocks to facilitate the transition between different levels of abstraction and to control the growth of feature maps. These transition layers typically include pooling operations, such as max pooling, which reduce the spatial dimensions of the feature maps and preserve important features at the same time. By reducing the size of feature maps, transition layers help manage computational complexity and prevent overfitting. This design choice enables DenseNet to achieve a balance between model capacity and computational efficiency. For a comprehensive understanding of the DenseNet network structure, refer to Figure 4.

### 4.2 Network Improvement

In traditional convolutional neural networks, the convolutional transformation, where the input is transformed into the output, involves summation across channels based on the convolution results of each channel. This process achieves feature fusion within a local spatial region. However, this approach overlooks feature fusion across channels, despite each channel carrying different levels of importance in feature information. Recognizing the limitation of traditional convolution in utilizing channel feature information, this study integrates an attention mechanism, specifically the Squeeze-and-Excitation (SE) module, into the DenseNet model. The structure of the SE module is depicted in Figure 5.

The SE module utilizes backpropagation to learn the weight coefficients for each feature channel, representing the importance of each channel. First, the squeeze operation aggregates global spatial information into a channel descriptor by employing global average pooling across spatial dimensions. This produces a channel-wise statistic—a vector whose elements represent global receptive fields for the corresponding channels. Second, the excitation operation involves learning a non-linear, channel specific gating mechanism. Utilizing the squeezed information, it employs a simple gating mechanism with a sigmoid activation to capture channel-wise dependencies. The resulting weights are employed to adaptively recalibrate the original feature maps by rescaling them with the learned activation Subsequently, it extracts feature information from channels based on these weight coefficients, enabling

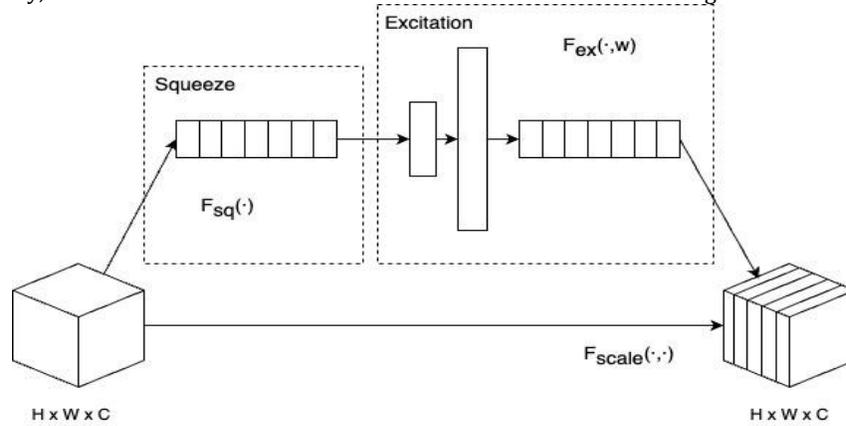

Figure 5. Squeeze-Excitation module.

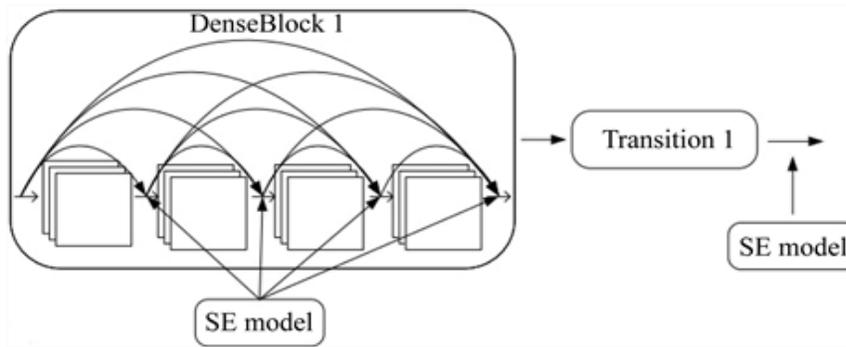

Figure 6. Schematic diagram of the attention mechanism embedding position.

feature fusion across channels and thereby enhancing network performance.

By introducing the squeeze-and-excitation (SE) operation on top of the DenseNet architecture, the network has been improved to achieve both spatial feature fusion and learning relationships between feature channels, further enhancing network performance.

The modified network structure, as depicted in Figure 6, incorporates the SE module into the DenseNet network's dense block sub-modules and behind the transition layers.

## 5 EXPERIMENTS

### 5.1 Evaluation Metrics and Experimental Environment
*5.1.1 Experimental Environment.*

The experimental setup includes a computer with the following specifications: 64-bit Windows 10 operating system, Intel Core i9-9900 CPU, 64 GB RAM, and NVIDIA GeForce RTX 4090 GPU. The experiments were conducted using the PyTorch deep learning framework.

Table 2. Related parameter settings.

| Parameter Name | Value |
| --- | --- |
| BatchSize | 32 |
| Epochs | 200 |
| Learning Rate | 0.01 |
| Dropout | 0.25 |

*5.1.2 Experimental Evaluation Metrics.*

In this experiment, the results are evaluated from the perspectives of patients and breast cancer pathological images, using classification accuracy as the evaluation metric. This helps analyze the classification performance of the proposed model and further improve it.

The patient-level classification accuracy is calculated as shown in Equation 1, and the average classification accuracy across all patients in the dataset is calculated as shown in Equation 2.

$$P_{rp} = \frac{N_{rp}}{N_{np}} \quad (1)$$

$$P_{arp} = \Sigma \frac{P_{rp}}{N_p} \quad (2)$$

Where:
$N_{np}$ represents the total number of a patient's pathological images.
$N_{rp}$ represents the number of correctly classified images for that patient.
$P_{rp}$ represents the classification accuracy of all pathological images for a single patient.
$P_{arp}$ represents the average classification accuracy across all patients in the dataset.
$N_p$ represents the total number of patients in the dataset. The classification accuracy from the perspective of breast cancer pathological images is calculated as shown in Equation 3:

$$P_{img} = \Sigma \frac{N_r}{N_{all}} \quad (3)$$

Where:
$P_{img}$ represents the classification accuracy of all pathological images.
$N_r$ represents the number of correctly classified pathological images.
$N_{all}$ represents the total number of images in the dataset after augmentation.

### 5.2 Experimental Strategy

The processed dataset is split into train, validation, and test sets in a ratio of 7:1:2 randomly. The training set is used for model training, including parameter learning; the validation set is used for model validation, continuously assessing the model's generalization ability and automatically fine-tuning parameters while saving the best model at any given time; the testing set is employed to evaluate the model's recognition rate and generalization ability. Additionally, all training data are shuffled before processing.

In the experiment, the setting of relevant parameters is crucial. Regarding the learning rate in deep learning, a value too large may result in exploding loss values, while a value too small may lead to overfitting, slowing down the convergence speed of the network model. Typically, an initial learning rate ranging from 0.1 to 0.001 is advisable. Thus, in this study, the initial learning rate is set to 0.01.

As for the setting of training batch size and number of iterations in the experiment, a batch size that is too small may impede the convergence of the network model, while a batch size that is too large may trap the network model in local optima. Moreover, increasing the batch size will require more training epochs to achieve the same accuracy. Taking these factors into account, the training batch size in this study is set to 32, and the number of iterations is set to 200. Overall, the relevant parameter settings in this study's experiment are presented in Table 2.

In this study, the training will focus on a binary classification task: distinguishing between benign and malignant breast cancer pathological images.

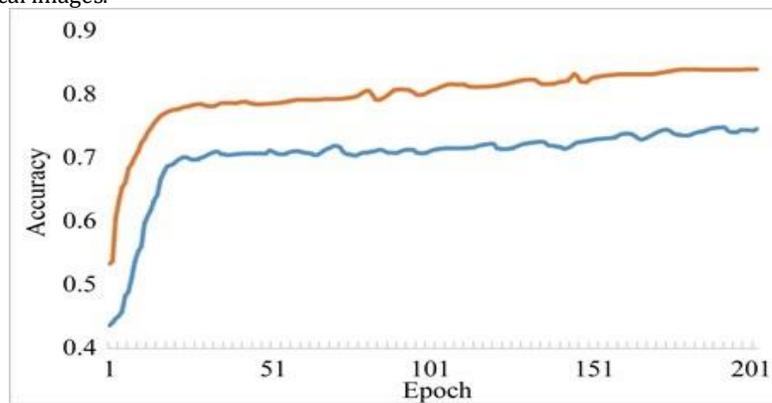

Figure 7. The training accuracy and the validation accuracy of the module.

### 5.3 Experimental Results and Analysis

The proposed classification model based on transfer learning and deep learning was continuously trained and optimized using the training and validation sets. The accuracy of the model on the training and validation sets is depicted in Figure 7. From the graph, it can be observed that the proposed classification model achieved a peak training accuracy of 82% as the red line shows and a validation accuracy of 75% as the blue line shows, with a difference of 7 percentage points between them.

Typically, without data preprocessing and augmentation, training accuracy tends to be high while validation accuracy may remain low and exhibit unstable fluctuations. However, through experimentation, it was demonstrated that preprocessing and augmenting the breast cancer pathological images can mitigate overfitting and enhance classification accuracy.

In this study, three network models were used for comparative validation experiments on breast cancer pathological images at different magnification levels. The three networks are DenseNet, DenseNet + SE, and the classification model proposed in this study based on transfer learning and deep learning. These models were evaluated for binary classification tasks on the BreakHis dataset at magnifications of 40×, 100×, 200×, and 400×, as shown in Table 3.

Table 3. Classification results of the three network models.

| Eavluation Metrics | Network | BreakHis Data Set | | | |
|---|---|---|---|---|---|
| | | 40× | 100× | 200× | 400× |
| Parp | DenseNet | 73.9 | 75.0 | 77.6 | 78.0 |
| | DenseNet + SE | 78.0 | 78.1 | 78.5 | 78.7 |
| | Ours | 80.1 | 84.3 | 81.2 | 82.4 |
| Pimg | DenseNet | 72.5 | 77.5 | 77.2 | 77.5 |
| | DenseNet + SE | 72.5 | 75.6 | 74.9 | 80.3 |
| | Ours | 78.4 | 79.2 | 79.7 | 84.0 |

It can be observed that the classification model proposed in this study, based on transfer learning and deep learning, shows an improvement in classification performance compared to the original network model, both from the perspective of patients and image analysis, with an increase of 2% to 6%.

Additionally, taking the magnification level of 400× as an example, the classification accuracies of the three network models are illustrated in Figure 8. In this study, the proposed classification model incorporates transfer learning on top of DenseNet + SE. As depicted in the figure, transfer learning provides the model with robust feature extraction capabilities as the green line shows, which contribute to the enhancement of classification accuracy.

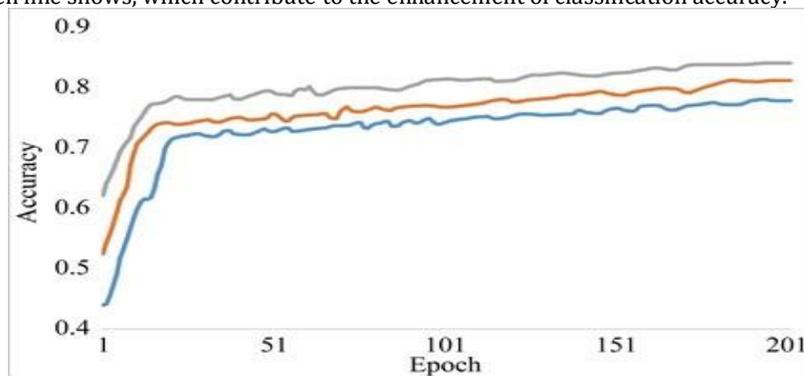

Figure 8. Experimental classification accuracy at 400× magnification.

Table 4. Number of parameters and model size of the network models.

| Network | DenseNet | DenseNet+SE | Ours |
|---|---|---|---|
| Parameter Number | 7.127k | 8.234k | 8.064k |
| Model Size | 82.4Mb | 89.5Mb | 84.7Mb |

The parameter counts and model size of the three network models are presented in Table 4.

From Table 4, it can be observed that compared to DenseNet and DenseNet + SE, our model possesses a slightly greater number of parameters and a larger overall model size. However, this increase in parameters and size is acceptable considering the improvement achieved in breast cancer classification.

As shown in Table 5 above, the convergence time of the three network models indicates that transfer learning is an effective strategy to address the problem with limited training data. By leveraging pretrained models, we can enhance the training efficiency and the network's ability to generalize. Without pre-training the weights of the network on a large dataset, the initial weights would be randomly set, leading to slower convergence of the network. In this study, we employed transfer learning to initialize the weights and fine-tuned them based on the images, thereby accelerating the convergence speed of the model during training.

## 6 CONCLUSION

This paper proposes a medical image classification method based on transfer learning and deep learning, targeting the complexity and limited scale of medical pathology tissue images. The model categorizes breast cancer pathology images from the BreakHis dataset into benign and malignant classes. Experimental results demonstrate the effectiveness of combining transfer learning with deep learning, leading to improvements in classification compared to the baseline model. However, the study has limitations. Firstly, the proposed model only performs binary classification of benign and malignant breast pathology tissues, without distinguishing the grading or subtyping of breast cancer. Additionally, the model parameters are slightly higher, and the model size is larger, without optimization in this regard. This highlights areas for future research and optimization. And above the accuracy, according to D, Ma et al.[13] how to balancing accuracy and interpretability to develop deep learning models that both doctors and patients can trust will become the research focus of the industry in the future.

Table 5. Convergence times of the network model.

| Network | Convergence time (minute) |
| --- | --- |
| DenseNet | 1237 |
| DenseNet + SE | 1356 |
| Ours | 976 |